# Derivation of longitudinal Doppler shift equation between two moving bodies in a reference frame at rest using the particle property of photons

Masanori Sato

*Honda Electronics Co., Ltd.,*
*20 Oyamazuka, Oiwa-cho, Toyohashi, Aichi 441-3193, Japan*

**Abstract:** The equation of the Doppler shift of two bodies in inertial motion in a reference frame at rest (i.e., stationary reference frame) is derived. In this derivation, the wave-particle duality of photons in the theory of special relativity is considered. The dilation of time of a moving clock, which is derived from the Lorentz transformation that depends on the velocity of the moving clock in the reference frame at rest, and wavefront counting by geometrical drawing are used to describe the longitudinal Doppler shift. We show that the equation of the Doppler shift depends on the velocities of the two bodies in the reference frame at rest.

1. Introduction

Wave-particle duality has been discussed in quantum mechanics, however there is no discussion in the theory of special relativity. Sato [1] discussed the Michelson-Morley experiment in terms of single photons using de Broglie-Bohm picture: that is, the Michelson-Morley experiment showed an interference condition and did not show the simultaneous arrival of two photons (this is because there is only single photon). The Michelson-Morley experiment shows the wave property of photons: the wave property is assumed to be a nonlocal quantum potential in Bohm theory [2]. It is important to point out that wave-particle duality should be discussed in the theory of special relativity as well as in quantum mechanics. Most of the counterintuitive aspects in the theory of special relativity arise from wave-particle duality. If wave-particle duality in the theory of special relativity is discussed clearly, a more intuitive interpretation of the theory of special relativity is possible.

The Doppler shift shows the particle property of photons: that is, the interference of photons does not appear. It is rather difficult to distinguish the properties of a wave and particle; at this stage, the wave property is defined by the interference experiments and the particle property is defined by the experiments without interference. Thus, frequency counting is defined by not wave but particle properties. In a Doppler shift experiment, the photon is measured as a particle, which travels at the speed of light in vacuum regardless of the velocity of the light source.

The Fizeau experiment, in which a refractive index is affected by the flow of water (the speed of light in water is affected by the flow of water), is also an interference experiment; a single photon interference experiment is possible. The aberration of light is detected without interference; a photon is detected as a particle. Thus, experimental results should be discussed from the viewpoint of wave-particle duality in the theory of special relativity, which is summarized in **Table 1**. The Michelson-Morley experiment shows the wave property of photons; thus, the inertial motion of the earth was not detected because the wave property of a photon is nonlocal, and the interference condition does not depend on the velocity of the frame. The particle property of photons is local, which means that they travel at the speed of light; thus, the particle property is affected by the inertial motion of the experimental setup.

In this paper, using the particle property of photons, we show that the Doppler shift of light is derived from a geometrical drawing and the Lorentz transformation of the reference time. We derive the equation of the longitudinal Doppler shift between two bodies moving relative to each other in a reference frame at rest. Furthermore, we show how to find the reference frame at rest.

**Table 1** Wave particle duality in the theory of special relativity

| Wave property (interference) | Particle property |
|---|---|
| Michelson-Morley experiment | Doppler shift |
| Fizeau experiment | Aberration of light |

2. Relative motion of three bodies

**Figure 1** shows the reference frame at rest and two inertial moving objects: O, the reference frame at rest, A, a moving object (rocket A) with the absolute velocity $v_A$, and B, a moving object (rocket B) with the absolute velocity $v_B$. Rockets A and B each have an atomic clock and a light source that are precisely adjusted with respect to the reference frame at rest. In this paper, an absolute velocity defined in the reference frame at rest is adopted, which will be discussed in section 5.

The absolute velocities $v_A$ and $v_B$ are defined in the reference frame at rest. Let the reference time in the reference frame at rest be $t_0$; the reference time in a moving frame is defined by the Lorentz transformation using $t_0$. The reference time of object A is $t_A$ and that of object B is $t_B$:

$$t_A = \frac{t_0}{\sqrt{1-\left(\frac{v_A}{c}\right)^2}} \quad , \tag{1}$$

$$t_B = \frac{t_0}{\sqrt{1-\left(\frac{v_B}{c}\right)^2}} \tag{2}$$

where $t_A$ and $t_B$ are the reference times of the moving objects seen from the reference frame at rest (that is, the dilation of time). We cannot measure $t_A$ or $t_B$ directly; rather, it can be detected through the Doppler shift frequency.

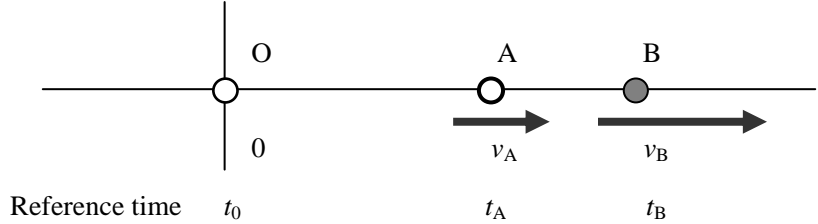

**Fig. 1** Doppler shift of moving bodies in a reference frame at rest

    O: Reference frame at rest                            $t_0$ (reference time)

    A: Moving object A (rocket A)                  $t_A$ (reference time)

    B: Moving object B (rocket B)                  $t_B$ (reference time)

3. Derivation of the longitudinal Doppler shift of light

   The longitudinal Doppler shift of light consists of the classical Doppler shift of sound (i.e., geometrical drawing) and the Lorentz transformation. **Figure 2** shows the way in which the longitudinal Doppler shift is derived; the classical Doppler shift determined from the geometrical drawing and the Lorentz transformation are combined. The geometrical drawing shows the counting of wavefronts. In **Fig. 2**, observer A counts the wavefronts of light from light source B. Thereafter, the Lorentz transformation is applied. The frequency of light from source B and the reference time of observer A with which the frequency is counted are modified by the Lorentz transformation.

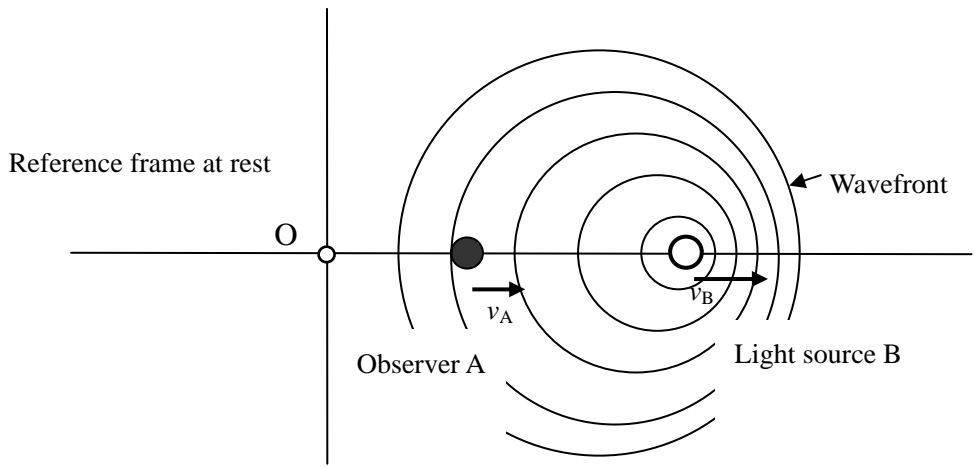

**Fig. 2** Derivation of the longitudinal Doppler shift from a geometrical drawing and the Lorentz transformation

   The classical Doppler shift of a frequency $f'$, represented by equation (3), is derived from a geometrical drawing, where $v_0$ is the velocity of the observer, $v_S$ is that of the light source, and $f_0$ is the frequency of the light source in a stationary state.

$$f' = f_0 \left( \frac{1 \pm v_0/c}{1 \pm v_S/c} \right) \tag{3}$$

here, + and – indicate the direction of the relative motion of the light source and the observer. The + or – sign is selected according to the relative motion between the light source and the detector. The frequencies are summarized in **Table 2**.

**Table 2**  Summary of frequencies

$f'$: Frequency of classical Doppler shift derived using geometrical drawing

$f_M$: Frequency of moving light source (inverse of reference time)

$f_0$: Frequency of stationary light source

$f_{A \to B}^L$: Longitudinal Doppler shift (object B is seen from viewpoint of object A)

$f_{B \to A}^L$: Longitudinal Doppler shift (object A is seen from viewpoint of object B)

In **Figs. 1** and **2**, when the observer at rest ($v_0$=0) on earth sees rocket B (the light source) leaving, then the + sign is selected to decrease the frequency according to $v_S$. Where $v_S$ is replaced by $v_B$, thus, equation (4) is derived from equation (3) as

$$f' = f_M \left( \frac{1}{1 + v_B/c} \right). \tag{4}$$

Then, we apply the Lorentz transformation. The frequency of the moving light source $f_M$ is derived as the inverse of the reference time $t_B$ of equation (1), as shown in

$$f_M = \frac{1}{t_B} = f_0 \sqrt{1 - \left(\frac{v_B}{c}\right)^2}. \tag{5}$$

By introducing equation (5) into $f_M$ in equation (4), we obtain

$$f_{O \to B}^L = f_0 \sqrt{\frac{1 - v_B/c}{1 + v_B/c}}. \tag{6}$$

Equation (6) is the Doppler shift derived from the geometrical drawing and Lorentz transformation. The stationary observer sees the light source moving at the velocity $v_B$.

When the moving observer with velocity $v_B$ sees the light source at rest, the geometrical drawing is obtained by setting $v_S$=0 in equation (3), and considering the direction of movement of the observer, we select –. Thus, we obtain

$$f' = f_0 \left(1 - \frac{v_B}{c}\right). \tag{7}$$

The reference time of the moving observer becomes long according to the Lorentz transformation. Thus, the frequency of the light source that is seen by the moving observer increases. This is because the reference time of the moving observer decreases. Equation (7) is modified as

$$f_{B \to O}^{L} = \frac{f_0}{\sqrt{1 - \left(\frac{v_B}{c}\right)^2}} \left(1 - \frac{v_B}{c}\right) = f_0 \sqrt{\frac{1 - v_B/c}{1 + v_B/c}}. \tag{8}$$

With equations (6) and (8), we show that the equation of the Doppler shift is equivalent for the observer and the light source. This is the case when either the observer or the light source is stationary.

In equations (6) and (8), the sign of the velocity, + or –, should be considered; a + velocity denotes that the distance between the observer and the light source increases, and a – velocity denotes that the distance between the observer and the light source decreases.

4. Derivation of the Doppler shift equation of two moving bodies in the reference frame at rest

Here, we discuss the relative motion of the two moving bodies in the reference frame at rest. The Doppler shift equations for rocket A and rocket B are derived. We see the light source in rocket B from the viewpoint of rocket A using the clock in rocket A as a reference. The observer is rocket A and the light source is rocket B, and the directions of the velocities of rockets A and B are chosen as shown in **Figs. 1** and **2**.

Therefore, the observer travels away from the light source and, at the same time, the light source travels towards the observer; thus, we choose + in equation (3). If the observer is rocket A and the light source is rocket B, then $v_0$ is replased by $v_A$ and $v_S$ is replased by $v_B$ and, thus, we obtain

$$f' = f_0 \left(\frac{1 + v_A/c}{1 + v_B/c}\right). \tag{9}$$

Then, we carry out time conversion, that is, time modification using the reference times of rockets A and B. The reference times of rockets A and B are modified using the Lorentz transformation. When the observer in rocket A sees the light source in rocket B, the reference time of rocket A in equation (1) is used.

The frequency of the light source is the inverse of equation (2). Thus, when the observer in rocket A sees the light source in rocket B, the correction constant for the Doppler shift is $t_A/t_B$. Multiplying equation (9) by $t_A/t_B$ we obtain

$$f_{A \to B}^{L} = f_0 \frac{t_A}{t_B}\left(\frac{1+v_A/c}{1+v_B/c}\right) = f_0 \frac{\sqrt{1-(\frac{v_B}{c})^2}}{\sqrt{1-(\frac{v_A}{c})^2}}\left(\frac{1+\frac{v_A}{c}}{1+\frac{v_B}{c}}\right) = f_0 \sqrt{\frac{1-\frac{v_B}{c}}{1-\frac{v_A}{c}} \times \frac{1+\frac{v_A}{c}}{1+\frac{v_B}{c}}}, \quad (10)$$

where $f_{A \to B}^{L}$ indicates the frequency at which the observer in rocket A sees the light source in rocket B. Equation (10) depends on $v_A$ and $v_B$, not on the relative velocity $v = v_A - v_B$. As discuss in section 5.1, if the both velocities $v_A$ and $v_B$ are not zero, the representation of the relative velocity $v = v_A - v_B$ is not correct: the relativistic velocity addition law should be adopted.

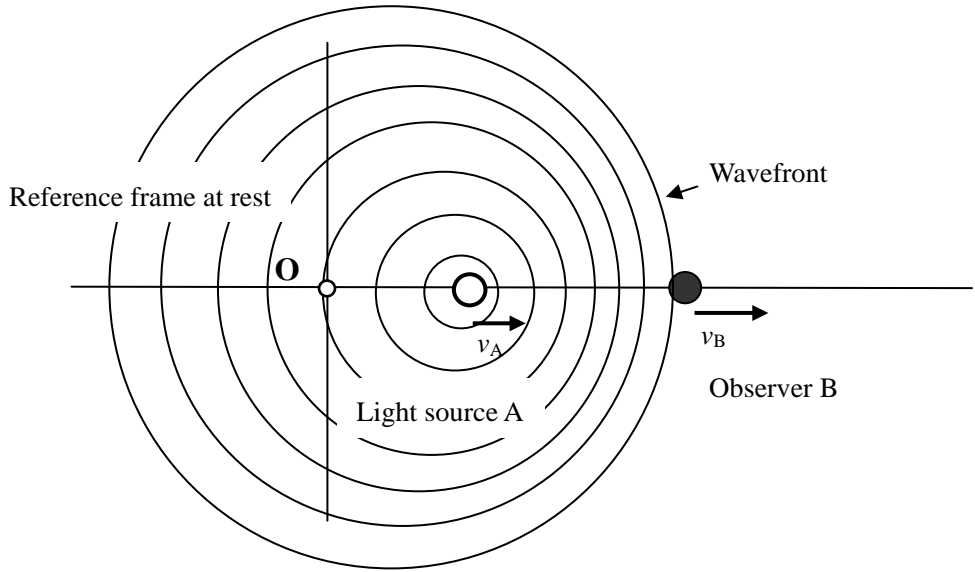

**Fig. 3** Derivation of longitudinal Doppler shift from geometrical drawing and Lorentz transformation. Observer and light source are changed from those in **Fig 2**.

**Figure 3** shows the observer in rocket B seeing the light source in rocket A, i.e., the observer is rocket B and the light source is rocket A. Thus, the observer moves away from the light source and the light source moves towards the observer. When we replace $v_0$ by $v_B$ and $v_S$ by $v_A$, and carefully choose + and – in equation (3), we obtain

$$f' = f_0 \left(\frac{1 \pm v_0/c}{1 \pm v_S/c}\right) = f_0 \left(\frac{1 - v_B/c}{1 - v_A/c}\right). \quad (11)$$

Then,

$$f_{B\to A}^L = f_0 \frac{t_B}{t_A}\left(\frac{1-v_B/c}{1-v_A/c}\right) = f_0 \frac{\sqrt{1-(\frac{v_A}{c})^2}}{\sqrt{1-(\frac{v_B}{c})^2}}\left(\frac{1-\frac{v_B}{c}}{1-\frac{v_A}{c}}\right) = f_0 \sqrt{\frac{1-\frac{v_B}{c}}{1-\frac{v_A}{c}} \times \frac{1+\frac{v_A}{c}}{1+\frac{v_B}{c}}}. \qquad (12)$$

From equations (10) and (12), we obtain

$$f_{A\to B}^L = f_{B\to A}^L. \qquad (13)$$

We note that the longitudinal Doppler shift depends on the absolute velocities of the observer and the light source rather than the relative velocity of the observer and the light source.

We consider equation (10); if we set $v_A = 0$ we obtain the longitudinal Doppler shift of equation (6). Under the condition $v_A = v_B$ (that is, the relative velocity is zero), the observers in both rockets see the same frequency $f_0$, that is,

$$f_{A\to B}^L = f_{B\to A}^L = f_0. \qquad (14)$$

Equation (10) shows not only the generalization of equation (6) but also the necessity of the absolute velocity. This paper points out the necessity of the absolute velocity in the Doppler shift equation.

5. Discussion
5.1 Representation using relativistic velocity addition law

Instead of the relative velocity $v = v_B - v_A$, if the relativistic velocity addition law is used as follows, (two velocities are $v_B$ and $-v_A$).

$$u = \frac{v_B + (-v_A)}{1 + \frac{(-v_A)v_B}{c^2}} = \frac{v_B - v_A}{1 - \frac{v_A v_B}{c^2}}, \qquad (15)$$

equations (10) and (12) are represented as

$$f_{A\to B}^L = f_{B\to A}^L = f_0 \sqrt{\frac{\left(1-\frac{v_B}{c}\right)\left(1+\frac{v_A}{c}\right)}{\left(1-\frac{v_A}{c}\right)\left(1+\frac{v_B}{c}\right)}} = f_0 \sqrt{\frac{1 - \frac{v_B - v_A}{c(1-\frac{v_A v_B}{c^2})}}{1 + \frac{v_B - v_A}{c(1-\frac{v_A v_B}{c^2})}}} = f_0 \sqrt{\frac{1-\frac{u}{c}}{1+\frac{u}{c}}}. \qquad (16)$$

The velocity $u$ in equation (16) contains absolute velocities $v_A$ and $v_B$ as represented in equation (15). However, equation (16) is in good agreement with the orthodox representation of the Doppler shift frequency. If one of the velocity $v_A$ or $v_B$ is zero, equations (6) or (8) can be derived. The postulate of an absolute reference frame is compatible with the essence of the Doppler shift representation in the theory of special relativity.

5.2 Application of the Doppler shift equation

We use the Doppler shift which counts wavefronts directly. The interference of the wave is not used, that is, the particle property of wave-particle duality is taken into consideration.

**Figure 4** shows the light source S moving with an absolute velocity of 30 km/s, rocket A at 34 km/s, and rocket A' at 26 km/s. According to equation (13), we obtain as follows,

$$f^L_{S \to A} = f^L_{A \to S}, \quad f^L_{S \to A'} = f^L_{A' \to S} \quad \text{and, generally,} \quad f^L_{S \to A} \neq f^L_{S \to A'}.$$

From **Fig. 4**, rocket A' counts more wavefronts than rocket A does; it is easy to predict that $f^L_{S \to A'} \geq f^L_{S \to A}$. Substituting c=300,000 km/s (the speed of light), $v_S$=30 km/s, $v_A$=34 km/s, and $v_{A'}$=26 km/s into equation (10), we obtain,

$$f^L_{S \to A} = f_0 \sqrt{\frac{1 - \frac{v_A}{c}}{1 - \frac{v_S}{c}} \times \frac{1 + \frac{v_S}{c}}{1 + \frac{v_A}{c}}} = 1.0001130 \times f_0, \tag{17}$$

and $f^L_{S \to A'} = 1.0001140 \times f_0$. Thus, the difference between the Doppler shift frequencies is

$$\frac{f^L_{S \to A'} - f^L_{S \to A}}{f_0} \approx 1.0 \times 10^{-7}. \tag{18}$$

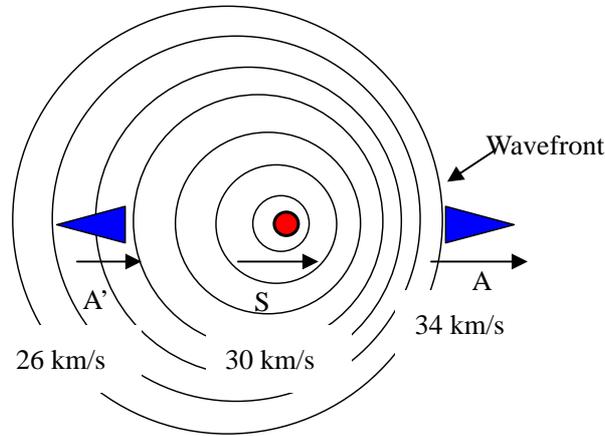

**Fig. 4** Relative velocities represented using equation $v = v_B - v_A$, however this illustration does not show the correct relative velocities. The correct relative velocity should be calculated using the equation of Doppler shift or equation (15).

The discussion of Doppler shift which uses the relative velocity represented as $v = v_B - v_A$ is not correct. We cannot illustrate the relative velocity as shown in **Fig. 4**. As discussed in section 5.1, the relative velocity should be represented as equation (15), that is, if we set $v_S$=30 km/s, $v_A$=34 km/s, the relative velocity is not calculated exactly 4 km/s but 4.000000045 km/s, and $v_{A'}$ is not exactly 26 km/s but 25.999999989 km/s. The relative velocity should be defined by the Doppler frequency. If we detect $f^L_{S \to A} = f^L_{S \to A'}$, the relative velocities between rocket S and rockets A and A' are equal.

5.3 To find the reference frame at rest by the sing around method

From the discussion in 5.2, we find that the relative velocity can be experimentally detected using Doppler shift. In this section, a simple method which can be used to find the reference frame at rest is described. Instead of the Doppler shift, the sing around method, which was previously tried by Galileo using lanterns on top of two mountains, is applied. In the case of acoustic waves, for the measurement of sound speed, the sing around method is used. The sing around experimental setup using a light source [1, 3], which uses two pairs of light sources and detectors as shown in **Fig. 5**, can be constructed, where a pulsed signal (a flash of light) is transmitted by light source 1 and detected by detector 2. After its detection, a new pulsed signal is transmitted by light source 2, detected by detector 1, transmitted again by light source 1, and so on. Light sources are constructed using LEDs, and the detectors are by photo diodes.

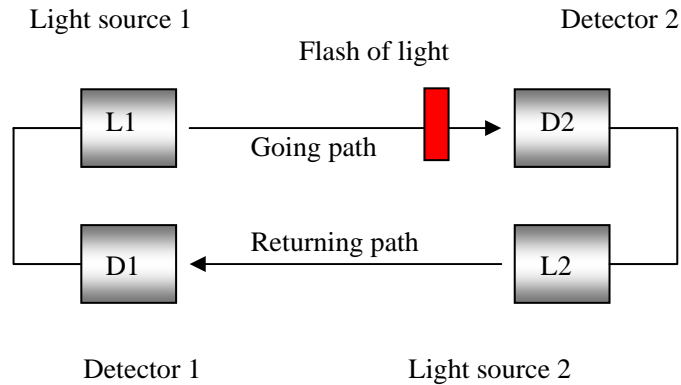

**Fig. 5** Sing around experimental setup using light

A flash of light from light source 1 is detected by detector 2. After its detection, a new flash of light is transmitted by light source 2, detected by detector 1, transmitted again by light source 1, and so on. This experiment is modern version of Galileo's experiment using lanterns on top of two mountains.

If we use the value shown in **Fig. 4** the differences are very small. We assume that the two rockets are traveling at the speeds of 40% (rocket S) and 60% (rocket A) of the speed of light c as shown in **Fig. 6**. According to equation (15) we obtain $u = \frac{5}{19}c$. Rocket S transmits a flash of light to rockets A and A' then they flash back to rocket S.

We calculate $v_{A'}$ using the equation of Doppler shift frequency (10) or equation (15). We obtain $v_{A'} = \frac{13}{85}c$ so that $f_{S \to A}^L = f_{S \to A'}^L$.

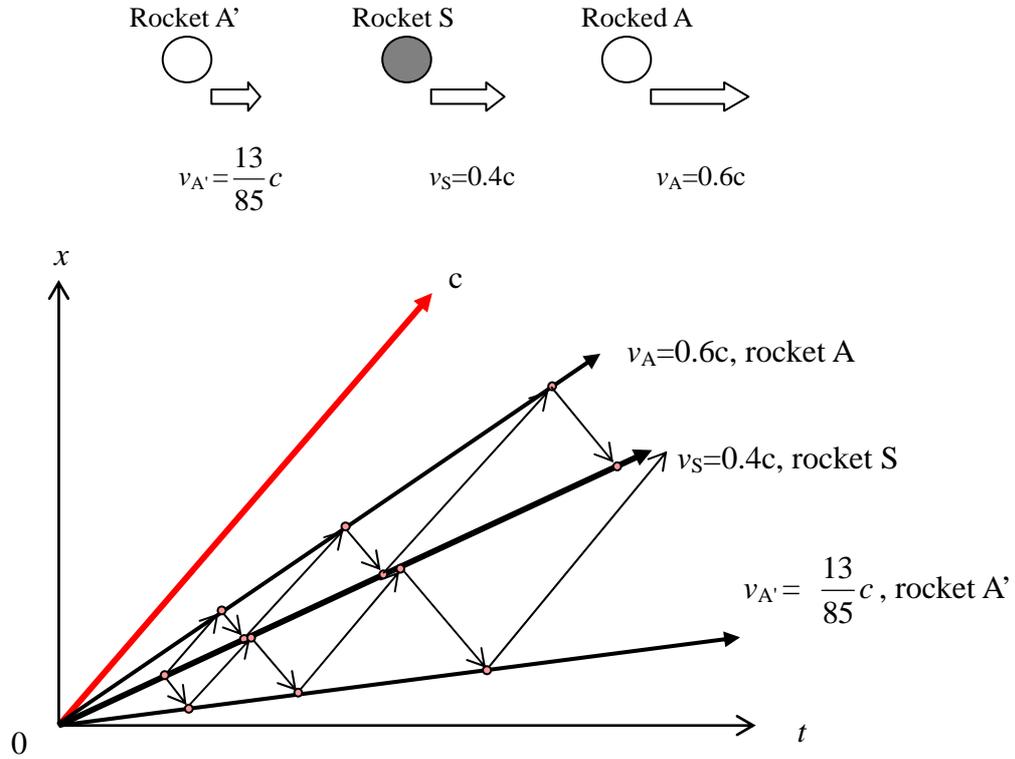

**Fig. 6** Sing around method of a light source: rockets A' and A are launched from rocket S in opposite direction in order to detect the same Doppler frequency. The speed of light, c is assumed to be constant regardless of the velocity of the light source. The light paths are drawn in the figure. The Doppler frequencies of rockets A and A' shown from rocket S is the same represented as $f_{S \to A}^L = f_{S \to A'}^L$. The relative velocities between rocket S and rockets A and B are equal. However, the arrival time of the return pulses detected at rocket S gradually sift. This indicates that rocket S is not in the reference frame at rest, and rocket S should be decelerated.

In this discussion, only Einstein's assumption, in which a photon travels at the speed of light, c, in vacuum regardless of the velocity of the light source, is used; thus, the three rockets will detect the flashes of light as shown in **Fig. 6**. If rockets A and A' are launched in opposite directions from rocket S, the flash pattern of light that rocket S detects is shown as the points on the line $v=0.4c$ in **Fig. 6**. Thus, we know the drifting direction of the rocket S in the reference frame at rest. Therefore, we can know how to decelerate rocket S to be close to the reference frame at rest.

The sing around method does not need the Lorentz transformation of the reference time: the dilation of time is not taken into consideration. It requires only a geometrical drawing; therefore, the discussion becomes clear and simple.

We propose a new method that is different from the Michelson-Morley experiment to detect the motion in the reference frame at rest.

Experimental procedure to find the reference frame at rest is:

(1) Rockets are started in all directions from rocket S and accelerated so as to detect the same Doppler shift frequency: $f_{S \to A}^L = f_{S \to A'}^L$, if the same Doppler shift frequency is detected all rockets are in the same relative velocity to rocket S.
(2) Detect the time interval of sing around pulses. Rockets S is moving toward the rocket which shows the shortest time interval of sing-around light pulses. Then, decelerate rocket S.
(3) Repeat the procedure (1) and (2) so as to detect the same time interval of sing around pulses for all rockets.

We have two methods to define the velocity: one is the Doppler shift frequency and the other is the repetition frequency of sing around light pluses. Thus we can detect the reference frame at rest. The merit of this representation is that the absolute reference frame can be detected. We know the absolute velocities of the two rockets so that there is no twin paradox.

6. Conclusion

The longitudinal Doppler shift of two moving bodies in a reference frame at rest was derived considering the particle property of photons in the wave-particle duality. The equation of the Doppler shift depends on the absolute velocities of the moving bodies. If the relativistic velocity addition law is used for the relative velocity between the observer and the light source, we can show that the longitudinal Doppler shift equation (16) is in good agreement with the orthodox theory of special relativity. The merit of this representation is that the absolute reference frame can be detected.

References

1) M. Sato, "Proposal of Michelson-Morley experiment via single photon interferometer: Interpretation of Michelson-Morley experimental results using de Broglie-Bohm picture," physics/0411217, (2004).
2) D. Bohm and B. Hiley, *The undivided universe*, (Routledge, London, 1993).
3) M. Sato, "Proposal of atomic clock in motion: Time in moving clock," physics/0411202, (2004).